\documentclass{aa}               % Option2 : Final A+A version (2 columns)
\input psfig.sty
\def\simless{\mathbin{\lower 1pt\hbox
   {$\spose{\raise 5pt\hbox{$\char'074$}}\char'430$}}}
\def\simgreat{\mathbin{\lower 1pt\hbox
   {$\spose{\raise 5pt\hbox{$\char'076$}}\char'430$}}}
\def\simgreat{\gapp}
\def\simless{\lapp}
\def\lapp{\mathbin{\raise2pt \hbox{$<$} \hskip-9pt \lower4pt \hbox{$\sim$}}}
\def\gapp{\mathbin{\raise2pt \hbox{$>$} \hskip-9pt \lower4pt \hbox{$\sim$}}}

\begin{document}

   \title{Nonradial and nonpolytropic  astrophysical outflows\protect\\
         VI. Overpressured winds and jets}
   \titlerunning{Nonradial and nonpolytropic  astrophysical outflows VI.}

  \author{C. Sauty
           \inst{1}
   \and    E. Trussoni
           \inst{2}
   \and    K. Tsinganos
           \inst{3}
          }

   \offprints{C. Sauty \\ (christophe.sauty@obspm.fr)}

   \institute
         {Universit\'e Paris 7, F\'ed\'eration APC 
-- Observatoire de Paris, LUTH, F-92190 Meudon, France
    \and Istituto Nazionale di Astrofisica (INAF) - Osservatorio Astronomico 
       di Torino, Strada Osservatorio 20, I-10025 Pino Torinese (TO), Italy
    \and  IASA and Section of Astrophysics, Astronomy \& Mechanics Department of Physics, University 
of Athens,
          Panepistimiopolis GR-157 84, Zografos, Greece
         }
   \date{Received 2 December 2003 / accepted 8 April 2004}

   \abstract{
By means of a nonlinear separation of the variables in the governing full set 
of the magnetohydrodynamic (MHD) equations for axisymmetric plasmas we analyse
an exact model for magnetized and rotating outflows which are hotter and 
overpressured at their axis. 
These outflows start subsonically and subAlfv\'enically from the central 
gravitating source and its surrounding accretion disk. Subsequently, they 
accelerate  thermally and magnetocentrifugally and thus cross the  
appropriate MHD critical points, reaching high values of the Alfv\'en Mach 
number. Three types of solutions are found :  
(a) collimated jet-type outflows from efficient magnetic rotators with the 
flow confined by the magnetic hoop stress; 
(b) radially expanding wind-type outflows analogous to the solar wind, 
from inefficient magnetic rotators or strongly overpressured sources; 
(c) terminated solutions with increasing amplitude of 
oscillations in the width of the beam. 
In contrast to previously studied underpressured outflows, the transition 
from  collimated jets to uncollimated winds is not continuous in the 
appropriate parametric space with a gap where no stationary solution is
found. Superfast at infinity solutions are filtered by three critical surfaces
corresponding to the three known limiting characteristics or separatrices 
of MHD wind theory.
Collimated and terminated solutions cross the slow, Alfv\'en and fast 
magneto-acoustic critical points. Radially expanding solutions cross the 
slow and Alfv\'en critical points while the last boundary condition is 
imposed by requiring that the pressure vanishes at infinity.
}

\maketitle


\section{Introduction}

A well known example which demonstrates analytically that 
astrophysical jets can be accelerated and collimated magnetically is the 
\cite{BP82}  model. 
This model has been shown to be actually the prototype of the wide family of 
the 
so-called {\it radially} self-similar disk wind-type outflows which has been 
recently reexamined analytically and numerically (e.g.  
\cite{OuyedPudritz97}, \cite{VT98}, \cite{Krasnopolskyetal99},
\cite{CasseFerreira00}, \cite{Ustyugovaetal00}, 
\cite{Krasnopolskyetal03}, { \cite{Kudohetal02}},
\cite{CasseKeppens03}).

A complementary wide class of MHD outflow solutions, which quantitatively 
demonstrated the transition of collimated outflows from efficient magnetic 
rotators to uncollimated outflows from less efficient magnetic rotators, is 
self-similar in the {\it meridional} direction (see \cite{STT02};  
henceforth STT02, and references therein). This class of models may describe
ordinary stellar winds, or collimated outflows composed of a central jet core
surrounded by a disk wind ({ Tsinganos \& Bogovalov, 2002}). 
Although this model is somewhat similar in geometry 
to an X-wind (e.g. \cite{Shuetal94,Shangetal02}), it nevertheless has some 
differences, such as that it consistently solves the full set of the MHD 
equations from the source to the far region and also that the connection 
between the disk and the magnetosphere is an X point rather than a
fan of concentrated magnetic flux. This class of analytical models may also 
be compared to the corresponding relaxation states of recent numerical 
simulations (e.g. {  \cite{Koideetal98}}, \cite{BT99},
  \cite{KeppensGoedbloed00}, { \cite{Mattetal03}, \cite{Koideetal00},
\cite{Koide03})  as is discussed in Sec. \ref{subsec6.2}}. 
 
In such meridionally self-similar models, one may either prescribe the 
poloidal 
structure of the streamlines, or assume a relationship between the radial and 
longitudinal components of the gas pressure gradient. The main properties 
of the first class of solutions which are asymptotically collimated are 
outlined in Trussoni et al. (1997; henceforth TTS97) wherein the essential role 
of rotation in getting cylindrical collimation has been demonstrated. On the 
other 
hand, if the two components of the pressure gradient are related, the
meridional structure of the streamlines is self-consistently deduced from 
the solution of the full set of the MHD equations.  Such rotating 
and magnetized outflows with a spherically symmetric structure of the 
gas pressure may be asymptotically superAlfv\'enic with radial or 
collimated fieldlines, depending on the efficiency of the magnetic 
rotator (\cite{ST94}; henceforth ST94).

In Sauty et al. (1999; henceforth STT99) we extended the results of ST94 by
performing an asymptotic analysis of the meridionally self-similar 
solutions for a non spherically symmetric structure of the
pressure. It was pointed out there that a superAlfv\'enic outflow may
encounter different asymptotic conditions where it can be thermally or
magnetically confined, and thermally or centrifugally supported.

Current-carrying {\it underpressured} flows with a pressure increasing 
as we move away from the axis, were studied in STT02. They were found to be 
either thermally or magnetically cylindrically collimated around their axis, 
depending on whether the efficiency of the magnetic 
rotator prevails or not to the thermal confinement, respectively.
They have been shown to be well suited to describe various astrophysical 
winds and jets (see \cite{Limaetal01, Meliani01, Sautyetal03}).

We complete here this work by studying {\it overpressured} outflows, i.e., with
a pressure decreasing away from the system axis. Such outflows can only be 
collimated via magnetic stresses if the magnetic rotator is sufficiently 
efficient. Otherwise the flow structure attains asymptotically a radial 
configuration. We present complete solutions that connect the base of the 
flow with its superAlfv\'enic regime. In particular we investigate if, 
and under which conditions, the basal region can be matched to the asymptotic
solutions outlined in STT99. Conversely to the previous study reported in 
STT02, the present analysis requires a very careful topological study of the 
MHD self-similar equations because of the presence of a second X-type 
magnetosonic critical point. 

In the following section 2 and in order to establish the used notation we 
briefly review the assumptions, parameters, variables and mathematical 
structure of the present model. In section 3 the asymptotic behaviour of the 
solutions presented in STT99 is also very briefly outlined. The results and 
parametric study are presented in Section 4 while the main properties of the 
three classes of cylindrical, radial and terminated solutions are summarized 
in section 5. 
Finally, in section 6 we discuss the astrophysical relevance of our results, 
in particular in relation to jets associated with young stellar objects.      

\begin{figure*}
%\vspace{11.0 cm}
\centerline{
\psfig{figure=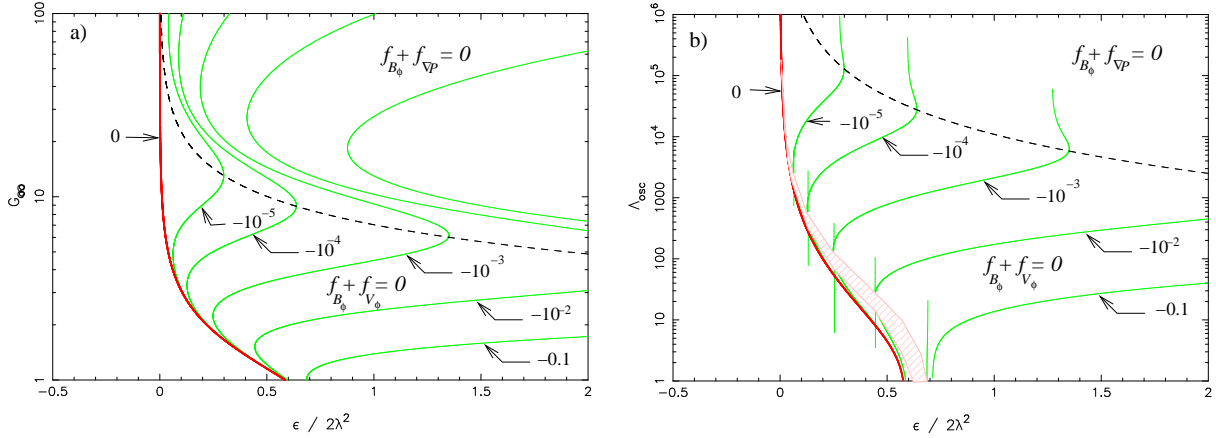,width=16.0truecm,angle=0}}
\caption{In a) is shown a plot of the dimensionless asymptotic 
radius of the jet $G_{\infty}$  
and in b) of the wavelength of the oscillations $\Lambda_{\rm osc.}$ 
in units of $r_*/\lambda$, vs. $\epsilon/2\lambda^2$ for a 
representative 
value of the terminal pressure, $\Pi_\infty=0.01$. Each curve is drawn for a 
constant value of $\kappa / 2\lambda^2$ which ranges from $-0.1$ to 
$-10^{-5}$. 
To the right of the dashed line is the domain of pressure supported and 
magnetically confined jets ($f_{B_{\phi}} + f_{\nabla P}$) while to the 
left of the dashed line is the domain 
of magnetocentrifugal jets ($f_{B_{\phi}} + f_{V_{\phi}}$) (STT99).
\label{f1}}
\end{figure*}

\section{Governing equations for meridional self-similar outflows}

We summarize here the main assumptions of our meridionally ($\theta-$)
self-similar treatment of the MHD equations. More details can be found in 
STT94, STT99 and STT02.

\subsection{Summary of the basic assumptions}

The basic equations governing plasma outflows in the framework of ideal MHD 
are the  momentum, mass and magnetic 
flux conservation equations, together with the frozen-in law for infinite 
conductivity and the first law of thermodynamics. First, with axisymmetry the 
poloidal 
component of the magnetic field can be derived from a magnetic flux function 
$A(r, \theta )$ in spherical coordinates ($r, \theta, \varphi$),
\begin{equation}\label{B}
\vec{B}= {\vec \nabla A \over r \, \sin \theta } \times \hat \varphi
\,.
\end{equation}

Next, for steady flows, we judiciously specify the meridional dependences of 
the velocity and 
magnetic fields, as well as of the density and pressure,  
$\vec{V}$, $\vec{B}$, $P$ and $\rho$ respectively, which may be written as 
follows 
(for details see ST94, STT99 and STT02):
\begin{equation}
\label{Br}
B_r = {B_{*}\over G^2(R)}\cos\theta\,,
\end{equation}
\begin{equation}
B_\theta =- {B_{*} \over G^2(R)}{F(R)\over 2}\sin\theta
\,,
\end{equation}
\begin{equation}
\label{Bphi}
B_\varphi = - {\lambda B_{*} \over G^2(R)}
{\displaystyle 1 - G^2(R) \over 1 - M^2(R) }{R\sin\theta}
\,,
\end{equation}
\begin{equation}
\label{Vr}
V_r = V_{*}  {M^2(R)\over G^2(R)} { \cos\theta \over
\sqrt{1+\delta \alpha(R,\theta)}  },\;\;\;\;\;\;
\end{equation}
\begin{equation}
V_\theta =-V_{*} {M^2(R)\over G^2(R)}{F(R)\over 2}
{  \sin\theta \over \sqrt{1+\delta \alpha(R,\theta)}  }
\,,
\end{equation}
\begin{equation}
\label{Vphi}
V_\varphi = {\lambda V_{*}  \over G^2(R)}
{ G^2(R) - M^2(R) \over 1- M^2(R)}
{R\sin\theta \over \sqrt{1+ \delta \alpha(R,\theta) } }
\,,
\end{equation}
\begin{equation}\label{density}
\label{rho}
\rho(R,\alpha) = {{\rho_*} \over {M^2(R)}} (1 + \delta \alpha)
\,,
\end{equation}
\begin{equation}\label{pressure}
P(R,\alpha) = {1 \over 2} \rho_* V^2_* \Pi(R)[1+ \kappa \alpha]
+P_o
\,.
\end{equation} 

In the above definitions, $\kappa$, $\lambda$, $\delta$ and $P_o$ are     
model parameters while we have also used the dimensionless magnetic 
flux function $\alpha(R, \theta) = 2 \, A(r, \theta) / r^2_* B_*$.   
Note that conversely to STT99 and papers before, we use a more flexible
definition of the pressure function, as in STT02. The dimensionless pressure 
along the polar axis is defined as $\Pi(R)$ within some free additive constant
$P_o$.  { It does not appear in the final dynamical equations as they
depend only on the pressure gradient. However, this constant can be adapted 
to the boundary conditions. In radial solutions where $\Pi$ goes
asymptotically to zero, $P_o$ should also vanish to ensure
that the temperature does not  diverge as the mass density also goes
asymptotically to zero. Conversely, in cylindrically collimated flows 
where $\Pi$ can be a negative function, $P_o$ should be adjusted such that the 
total pressure
remains  everywhere  positive.}

The square of the poloidal Alfv\'en number 
\begin{equation}\label{M}
M^2 \equiv M^2(R)= 4 \pi \rho {{V^2_p} \over {B^2_p}}
\,, 
\end{equation}
is assumed to be solely a function of the radial distance. 
For convenience we have normalized all quantities at the 
Alfv\'en surface along the rotation axis, $r=r_*$. The dimensionless radial 
distance is denoted by $R=r/r_*$, while $B_*$, $V_*$ and $\rho_*$ are the 
poloidal magnetic field, velocity and density along the polar axis at the 
Alfv\'en radius $r_*$, with $V^2_*= B^2_* / 4 \pi \rho_*$. 

A second assumption is that the cylindrical distance $\varpi (R, \alpha )$ of 
a poloidal fieldline from the axis is separable in the variables $R$ and  
$\alpha$, as $\varpi^2 = G^2 (R)\alpha$, where $G^2(R)$ is the cross sectional
area of a flux tube perpendicular to the symmetry axis, in units of the 
corresponding area at the Alfv\'en distance.  Then, the dimensionless magnetic
flux function $\alpha(R, \theta)$ is related to $G(R)$ through the following 
expression
\begin{equation}\label{alpha}
\alpha = {{R^2} \over G^2(R) } {\rm sin}^2 \theta
\,. 
\end{equation}
\noindent

Finally, for homogeneity with the notations in  ST94, STT99 and STT02,
we have also introduced the  function $F(R)$, which is the negative 
logarithmic derivative of the well known expansion factor used in 
solar wind theory (\cite{KoppHolzer76}):
\begin{equation}\label{F}
F(R) = 2 \, \left [1 -
{{\mathrm d} \, \ln G(R) \over {\mathrm d} \, \ln R} \right ]
\,.
\end{equation}
We recall that the value of $F$ defines the shape of the
poloidal streamlines. 
For $F(R) =  0$ the streamlines are radial, for $F(R) > 0$ they are 
deflected towards the polar axis (with $F=2$ corresponding to cylindrical 
collimation) while for $F(R) < 0$  they flare towards the equatorial plane.

\subsection{Parameters and variables \label{sec21}}

The model is controlled by the following four parameters.
\begin{itemize}

\item
The  parameter $\delta$ which governs the non spherically symmetric 
distribution of the density with a linear increase (or decrease) of the 
density when receding from the rotational axis for $\delta >0$ ($\delta <0$).

\item
The parameter $\lambda$ which is related to the rotation of the  poloidal
streamlines at the Alfv\'en surface $R=1$.

\item
The parameter $\kappa$ which controls the non spherically symmetric
distribution of the pressure. For $\kappa < 0$ ($\kappa > 0$) the gas pressure 
decreases (increases) by moving away from the polar axis. In this paper we 
confine our attention to overpressured jets, i.e., when $\kappa< 0$.

\item
The gravitational field is written as 
\begin{equation}
\vec{g} = - {{\cal G M}  \over r^2} \, \hat r
= -{1 \over 2} {V^2_* \over r_*}
{\nu^2 \over  R^2 } \, \hat r
\,,
\label{grv1}
\end{equation}
where ${\cal M}$ is the central gravitating mass. As a consequence there is an 
extra parameter $\nu$ which is the ratio of the escape and flow speeds  
at the Alfv\'en surface on the polar axis ($R=1$), 
\begin{equation}
\nu^2 ={{\cal G M}   \over r_* V^2_*}
\,.
\label{grv2}
\end{equation}
\end{itemize}

With the assumed axisymmetry, the original system of 
the MHD equations reduces to two coupled partial differential equations for 
the density and the magnetic flux. Furthermore, with the self-similarity 
assumption, the components of $\vec{V}$ and $\vec{B}$ can be written
as functions of $\theta$ and three functions of $R$, namely 
$G(R)$, $F(R)$ and $M(R)$. In this way, the momentum conservation 
law reduces to three ordinary differential equations which together with Eq. 
(\ref{F}) can be solved for the four variables $M^2(R)$, $F(R)$, $\Pi(R)$ and 
$G(R)$ (see Appendix A). 

\subsection{Efficiency of the magnetic rotator}

By integrating the momentum equation along a fieldline we obtain the 
conserved total energy flux density per unit of mass flux density. This is 
equal to the sum of the kinetic and gravitational energies, together with the 
enthalpy and net heating along a specific streamline. In the framework of the 
present meridionally self-similar model, the variation of the energy across  
poloidal fieldlines gives an important extra parameter (STT99):
\begin{eqnarray}
\epsilon  ={M^4\over (GR)^2}\left[ {F^2\over 4} - 1 \right]
- \kappa {M^4\over G^4}
- {(\delta\,- \kappa) \nu^2 \over R}
\nonumber\\
\label{EnEps}
+ {\lambda^2 \over G^2} \left({M^2-G^2 \over 1-M^2}\right)^2
+ 2\lambda^2{1-G^2 \over 1-M^2}\,,
\label{epsilon}
\end{eqnarray}
which is a constant on \underbar{all} streamlines (ST94).

Physically, $\epsilon$ is related to the \underbar{variation} across the 
fieldlines of the specific energy which is left available to collimate the 
outflow once the thermal content converted into kinetic energy and into 
balancing gravity has been subtracted (STT99). 

We can express $\epsilon / 2\lambda^2$  in terms of the conditions at the 
source boundary $r_o$ (see STT99 for details), 
\begin{equation}
{\epsilon  \over 2\lambda^2} =
{  E_{{\rm {Poynt.}},o} + E_{{\rm R},o} 
+\Delta E_{\rm G}^* 
\over E_{\rm {MR}}}
\,,
\end{equation}
where $E_{\rm {MR}}$ is the energy of the magnetic rotator (see Eq. 2.5a 
 in STT99), $E_{{\rm {Poynt.}},o}$
is the Poynting energy, $E_{{\rm R}_o}$ is the rotational energy at the base 
and $\Delta E_{\rm G}^*$ is the excess or deficit on a nonpolar streamline 
compared to the polar one of the gravitational energy (per unit mass) which is
not compensated by the thermal driving,
\begin{equation}
\Delta E_{\rm G}^* 
= - {{\cal G}{\cal M} \over r_o}
\left[  1-{T_o(\alpha)\over T_o({\rm{pole}})} \right]
=  -{{\cal G}{\cal M} \over r_o} {(\delta - \kappa )\alpha 
\over 1 + \delta \alpha }
\,.
\end{equation}

For $\epsilon >0$ collimation is mainly provided by magnetic means, while for 
$\epsilon <0$ the outflow can be confined only by the thermal pressure 
gradient, something which is 
not possible for overpressured flows. Accordingly, in STT99 we defined flows 
with positive or negative $\epsilon$ as {\it Efficient} or {\it Inefficient} 
Magnetic Rotators, respectively ({\bf EMR} or {\bf IMR}).

A solution is determined by the four parameters $\nu$, $\epsilon$, $\kappa$ 
and $\lambda$. The parameter $\delta$ can be deduced from the constraint 
imposed by the integral $\epsilon$, Eq. (\ref{EnEps}), which has the following 
expression at the Alfv\'enic singular surface ($R=1$):
\begin{equation}\label{epsilon2}
\epsilon = (\kappa - \delta) \nu^2  +\lambda^2 (\tau^2 + 1) - (1 -
\kappa) + F^2_*/4 \,,
\end{equation}
where $\tau [= (2 - F_*)/p]$ is given by Eq. (\ref{Eq8}).

\section{Asymptotic behaviour of the solutions\label{sec3}}

For $R \gg 1$ the asymptotic parameters of collimated outflows 
($F_{\infty} =2$, $G_{\infty}$ and $M_{\infty}$ bounded) depend on the value 
of $\epsilon$. Force balance across the poloidal streamlines, 
$f_{\nabla P} + f_{B_{\phi}} + f_{V_{\phi}}=0$, with  $f_{\nabla P}$, 
$f_{B_{\phi}}$ and $f_{V_{\phi}}$ the 
pressure gradient, magnetic stress and centrifugal volumetric force, 
respectively, 
calculates $M_{\infty}$ and $G_{\infty}$ as functions of the parameters 
$\epsilon/(2 \lambda^2)$, $\kappa/(2\lambda^2)$ and $\Pi_{\infty}$.

The asymptotic properties of these self-similar winds have been discussed in 
detail in STT99, and here we briefly summarize their main features for the 
case of overpressured outflows ($\kappa < 0$), some of which are displayed in 
Fig. \ref{f1}.

\begin{itemize}
\item{} Two main asymptotic regimes exist. In one the outflow is collimated by 
the pinching  of the toroidal magnetic field 
($\epsilon > \epsilon_{\rm lim}>0 $)
and in the other the outflow expands radially 
($\epsilon < \epsilon_{\rm lim}$).

\item{} For $\kappa \rightarrow 0$ we have collimation for any value of 
$\epsilon > 0$: there is no pressure gradient across the streamlines and the 
flow can be supported only by the centrifugal force 
($f_{B_{\phi}} + f_{V_{\phi}}=0$)

\item{} Magnetically collimated flows are supported either by the centrifugal 
force or by the thermal pressure. For a given set of values of $\kappa/2
\lambda^2$ and $\Pi_{\infty}$, solutions with an increasing asymptotic radius
$G_{\infty}$ are found to pass from centrifugally supported to pressure 
supported. For each regime and a given value of  $\epsilon/2 \lambda^2$ two 
solutions are found with a different asymptotic radius $G_{\infty}$.

\item{} Collimated streamlines always show oscillations. This behaviour is
consistent  with  the results  found in more general, non self-similar
treatments (\cite{VT98}) and numerical simulations. However as we move along a
 given curve with a fixed 
value of $\kappa/(2\lambda^2)$ and in the direction of increasing 
$G_{\infty}$ (see Fig. \ref{f1}), in the region of 
centrifugally supported flows, $\epsilon/(2 \lambda^2)$ decreases and then 
increases again. In the region of minimum $\epsilon/(2 \lambda^2)$ where the 
two regimes of centrifugally supported solutions merge we see that the 
wavelength becomes imaginary. This suggests that in this region
cylindrical asymptotics is unstable.

\end{itemize}

In conclusion and within the present model, from the asymptotic 
analysis it turns out that overpressured meridionally self-similar 
outflows from {\bf IMR} should always 
expand radially 
with an asymptotically vanishing pressure $\Pi_\infty=0$. Conversely,
outflows from {\bf EMR} should undergo a transition from 
radially expanding to cylindrically collimating, as the efficiency of the 
magnetic rotator increases.

\begin{figure*}
%\vspace{11.0 cm}
\centerline{
\psfig{figure=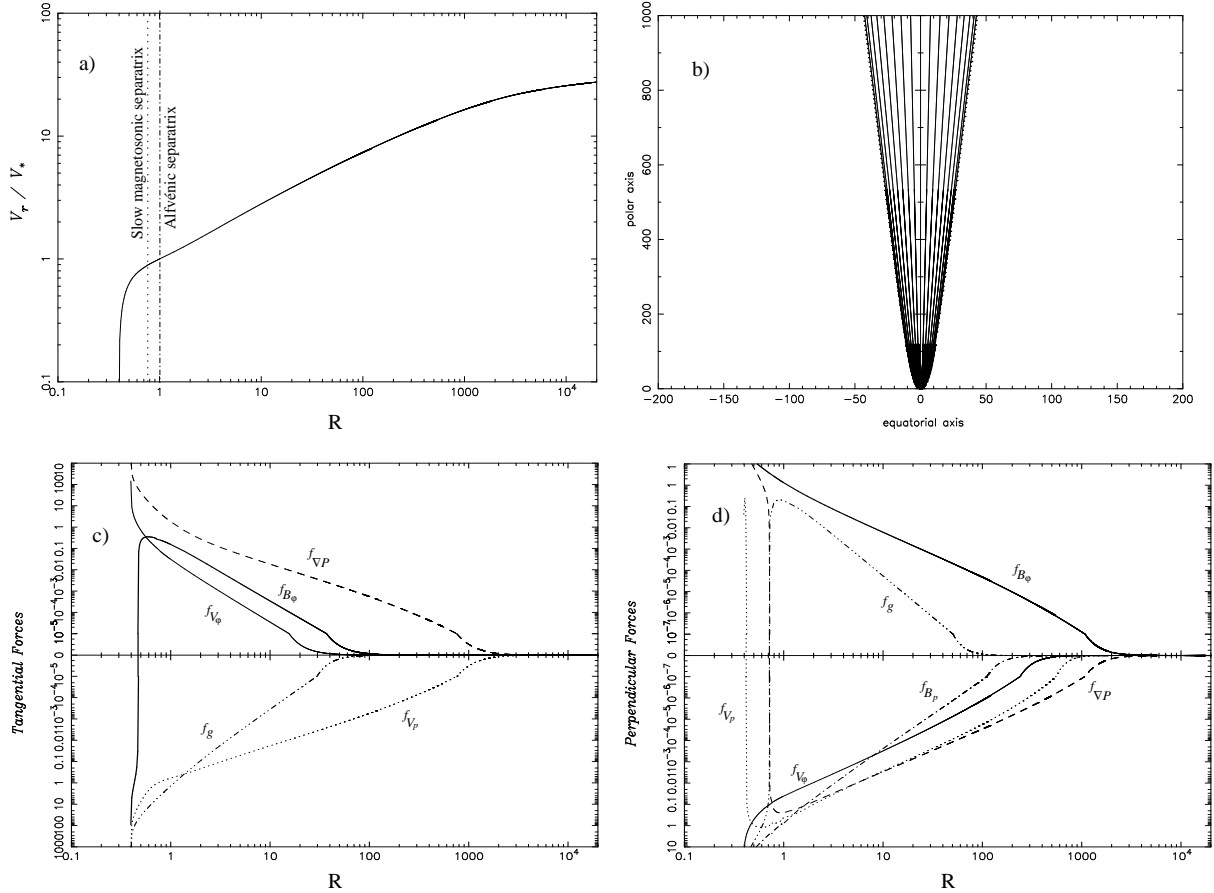,width=16.0truecm,angle=0}}
\caption{Typical example of a {\it radially} expanding solution. 
In a) $V_r/V_*$ along the polar axis is plotted vs. $R$ the radial
distance in units of the polar Alfv\'en radius. The dotted line 
corresponds to the slow critical point and the dot-dashed one to the Alfv\'en 
point. In b)  the shape of the poloidal streamlines is plotted for 
$\epsilon/(2 \lambda^2) 
= -0.05$ and $\kappa / (2 \lambda^2) = -0.0005$,  with the dotted 
line indicating a streamline which is not connected to the 
{ central star but to the surrounding accretion disk}.
The various volumetric forces acting along and perpendicular 
to a given streamline are also plotted vs. $R$ { in the two lower
  panels. In c) positive 
forces (upper half) acting along the flow are accelerating forces
while negative ones (lower  half) are decelerating ones. The 
inertial force  is negative as it is the opposite of acceleration. 
In d) positive forces (upper half) tend to collimate while negative
forces  (lower half) are decollimating. Symbols of the forces are defined in 
the text.}
\label{f2}}
\end{figure*}

\begin{figure*}
%\vspace{11.0 cm}
\centerline{
\psfig{figure=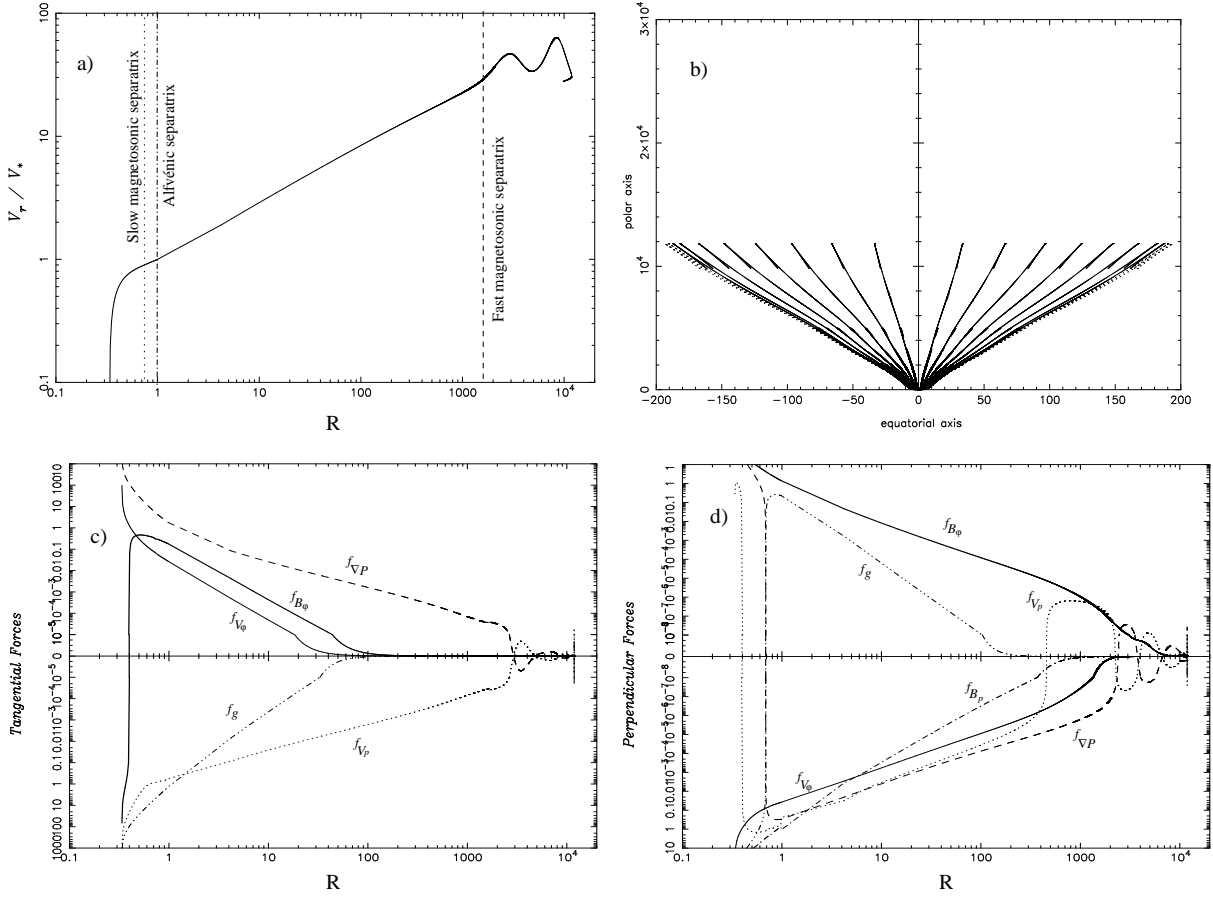,width=16.0truecm,angle=0}}
\caption{Typical example of a {\it terminated} critical solution.
The same plots to those of Fig. \ref{f2} are shown for  
$\epsilon/2 \lambda^2 = 0.05$ and  $\kappa/2 \lambda^2 = -0.00015$.
In a) the dotted line corresponds to the slow critical point, the 
dot-dashed one to the Alfv\'en point and the dashed line to the fast 
critical point. { Positive forces are accelerating in the upper part of c) 
and collimating in the upper part of d), while negative forces decelerate in 
c) and decollimate in d).}
\label{f3}}
\end{figure*}

\begin{figure*}
%\vspace{11.0 cm}
\centerline{
\psfig{figure=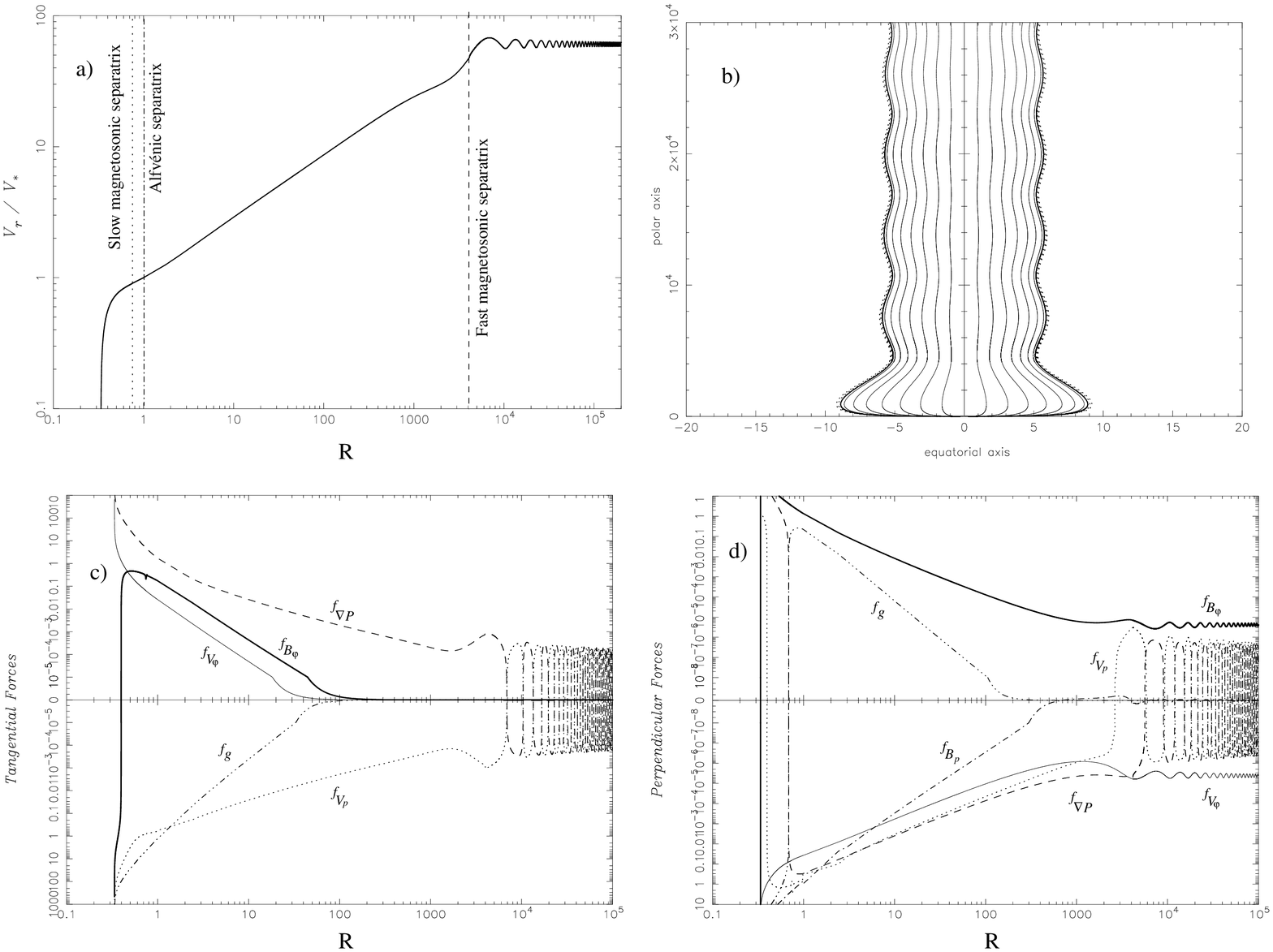,width=16.0truecm,angle=0}}
\caption{Typical example of a {\it collimated} critical solution. 
The same plots to those of Fig. \ref{f2} for $\epsilon/2 \lambda^2 = 0.05$ 
and  $\kappa/2 \lambda^2 = -0.000005$.
In a) the dotted line corresponds to the slow critical point (SMSS), the 
dot-dashed to the Alfv\'en point and the dashed to the fast 
critical point (FMSS). { Positive forces are accelerating in c) (upper part)
 and collimating in d) (upper part), while negative forces decelerate in c) 
and decollimate 
in d).}
\label{f4}}
\end{figure*}

\section{Numerical results\label{sec4}}

\subsection{Numerical technique}\label{subs41}
As in STT02, using routines of the NAG 
scientific package suitable for the  treatment of stiff systems and the
Runge-Kutta  algorithm, Eqs. (\ref{Eq1}) - 
(\ref{Eq7}) and (\ref{F}) are integrated upstream and downstream of the 
vicinity of the Alfv\'en transition ($R_{in} = 1 \pm {\rm d} R$) with 
$M_{in}=1 \pm p \, {\rm d}R$ and $G_{in} = 1 \pm (2 - F_{in}) {\rm d}R$ 
($F_{in} \approx F_*$). The slope $p$ of $M$ at $R=1$ is given in Eq. 
(\ref{Eq8}). We first integrate upstream tuning the value of $F_{in}$ until we
select the critical solution that smoothly crosses the singularity 
corresponding to the slow magnetosonic point and reaches the base of the wind 
$R_o$ with $M \rightarrow 0$. With this value of $F_{in}$ we then integrate 
downstream to the asymptotic region (with $R_{\infty}$ usually between $10^4$ 
and $10^6$).

Then, if the solution tends to 
become asymptotically radial or paraboloidal with a non zero pressure, we find
that the transverse gradient of the pressure  
dominates and forces the flow streamlines to
eventually flare towards the equator (or the pole if pressure is negative)
at a finite distance. Such a solution is terminated.
As is well known 
from the Parker wind theory, a physically acceptable solution which obtains 
radial asymptotics should satisfy the correct boundary condition at infinity, 
namely that the pressure should go to zero there. For this reason, we simply 
tune 
the value of the pressure $\Pi_{in}$ $(\approx \Pi_*)$ such that 
$\Pi (R\rightarrow \infty)$ vanishes. 
The various forces acting along and normal to a poloidal streamline 
are indicated in Figs. \ref{f2}, \ref{f3}, \ref{f4} as follows (see STT02): 
$f_g$, for the gravitational volumetric force, $f_{V_p}$ for the inertial 
volumetric force, $f_{B_p}$ for the poloidal magnetic volumetric force, and 
$f_{\nabla P}$, $f_{V_{\phi}}$, $f_{B_{\phi}}$ as defined in section 3.

If an extra critical point appears downstream of the Alfv\'en transition, we 
tune the value of the pressure $\Pi_{in}$ $(\approx \Pi_*)$ such that the 
solution crosses this second X-type critical point. After the critical point 
the solution is either collimated or terminated. We can always adjust the value
$P_o$ such that the total pressure remains positive everywhere (e.g. Figs. 
\ref{f3} and  \ref{f4}). 

Finally, if the solution naturally collimates and does not cross the second 
critical point, we obtain two alternatives. 
Either we choose $P_o=0$ as in STT99 and tune the 
value of the pressure $\Pi_{in}$ $(\approx \Pi_*)$ such that  $\Pi (R)$ is 
positive everywhere. Or, we let $P_o\neq 0$ as in STT02 (e.g. Figs. \ref{f4}),
and choose a value of $P_o>0$ such that the total pressure 
remains positive everywhere. 

It is worth to note that for $\kappa<0$, even though the pressure along the 
polar axis is always positive, it becomes negative for those nonpolar 
streamlines which correspond to $\alpha \ge {-1/\kappa}$, as it happened in 
TTS97. Thus, conversely to the $\kappa>0$ collimated solutions analysed in 
STT02, the present solutions cannot be extended to all streamlines away from 
the flow axis. This limitation however is expected since is well known that 
the meridionally self-similar solutions are more adapted to describe the flow 
close to its axis (cf. ST94) than far from it.

\subsection{Behaviour of the solutions with $\kappa$ and $\epsilon$}
\label{subs42}

In this subsection we have fixed $\nu=1 $ and $\lambda=1$ 
(with the value of $\delta$ deduced from Eq. \ref{epsilon2}) and analysed the 
trends of the solutions for different values of $\epsilon/2 \lambda^2$, 
$\kappa / 2 \lambda^2$ and $\Pi_*$. Even though this  
is a rather restricted set which does not exhaust the whole space of the 
parameters, it may nevertheless illustrate the main characteristics exhibited 
by the solutions.

\begin{figure*}
%\vspace{11.0 cm}
\centerline{
\psfig{figure=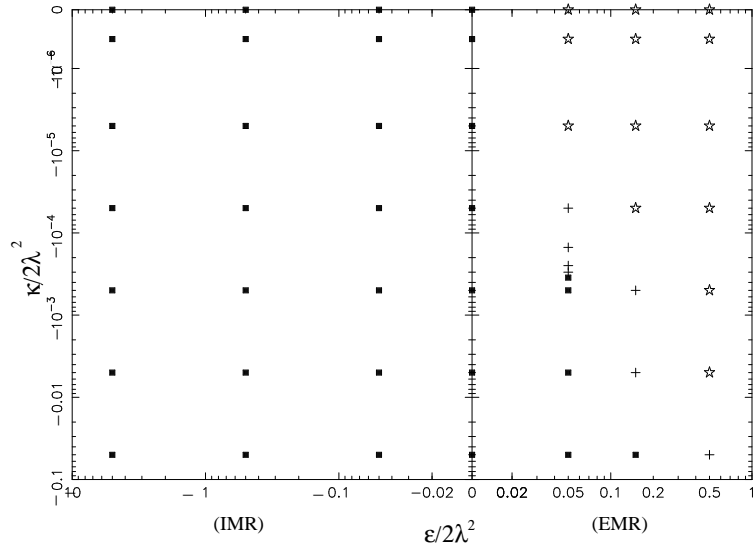,width=10.0truecm,angle=0}}
\caption{Regions of the radially expanding (full squares), terminated (crosses)
and collimated solutions (stars) in the plane of $\epsilon/2 \lambda^2$ and  
$\kappa/2 \lambda^2$.
\label{f5}}
\end{figure*}
 
For a constant negative value of  $\kappa$ and for $\epsilon$ increasing 
from negative to positive values three types of solutions are successively 
found. 
In the $\kappa/2 \lambda^2$ {\it vs.} $\epsilon/2 \lambda^2$ plane of 
Fig. \ref{f5} are sketched the three distinct asymptotic regimes of the 
outflow: 
\begin{itemize}
\item
{\it \underbar{Radial} asymptotics solutions}.  
An example of such a solution is shown in Figs. \ref{f2}, where we have 
plotted the dimensionless radial speed along the polar axis vs. the radial 
distance $R$, the shape of the poloidal streamlines and the volumetric forces 
acting along and perpendicular to the flow.  Such solutions correspond to the 
filled squares in the [$\epsilon/2 \lambda^2$, $\kappa / 2 \lambda^2$] plane 
displayed in Fig. \ref{f5}.
\item
{\it \underbar{Terminated} oscillating solutions}, with a typical example 
shown in Figs. \ref{f3}. 
Terminated solutions correspond to the crosses of the [$\epsilon/2 \lambda^2$, 
$\kappa / 2 \lambda^2$] plane (Fig. \ref{f5}).
\item 
{\it \underbar{Cylindrical} asymptotics solutions}, with a typical example 
shown in Figs. \ref{f4}. 
Asymptotically cylindrical solutions correspond to the stars of  the 
[$\epsilon/2 \lambda^2$, $\kappa / 2 \lambda^2$] plane (Fig. \ref{f5}).
\end{itemize}

The same evolutionary trend is found for a given positive value of $\epsilon$ 
by  increasing  $\kappa$ from negative values (see Fig. \ref{f5}). The 
character of the solutions changes from radial asymptotics to cylindrical ones
through a transition where the solutions are terminated.

If $\epsilon < 0$, we are in the regime of IMR and the transverse pressure 
gradient does not let the flow to cylindrically collimate. In fact, solutions 
with a nonvanishing pressure are either terminated with negative pressure, or 
they have excessive flaring ($F=-2$) such that all streamlines close at the 
equator, a rather unphysical situation as discussed in ST94.
A third type of solutions is found when the pressure at the Alfv\'en surface 
$\Pi_\star$ is tuned such that $\Pi_\infty$ vanishes, as we mentioned above. 
Then, as predicted by the asymptotic analysis, the flow streamlines 
asymptotically expand radially, with the Alfv\'en number increasing 
unboundedly far from the base while the flow speed is bounded 
($V_r\longrightarrow V_0, B_r \longrightarrow R^2, M \longrightarrow R$).
For a given $\epsilon$, by decreasing $\kappa$ to more negative values 
(Fig. \ref{f5}), the solutions with radial asymptotics have a decreasing 
terminal velocity and initial pressure $\Pi_{\star}$. This can be understood 
as follows.  The transverse pressure gradient is proportional to $\kappa\Pi$, 
and lower values of $\Pi_{\star}$ are needed to open the lines radially and at
the same time as the flaring is higher the velocity is lower as discussed in
\cite{TS92a}. 
This last result is unexpected from polytropic wind theory where a 
larger flaring leads to larger velocities \cite{KoppHolzer76} but this is 
precisely what has been observed for the fast component of the solar wind 
during the minimum and the maximum of the last solar cycles (see \cite{Wang95,
WangSheeley03}).

For  $\epsilon > 0$ and $\kappa$ lower than some threshold value $\kappa_1$, 
the same behaviour is observed. Solutions with an asymptotically vanishing 
pressure are radial. Other solutions are either flaring with $F_\infty=-2$ or 
refocalizing on the axis if the pressure becomes negative and then flaring 
again with $F_\infty=-2$.

When $\kappa$ reaches the value $\kappa_1$, we have the transition from 
squares to crosses in Fig. \ref{f5}. Now a second X-type critical point 
emerges in the superAlfv\'enic regime for $R > 1$.
Numerically, this second critical point appears at a finite distance and is 
not coming from infinity upstream, as one would expect.
We notice further that, assuming $|\kappa| \ll 1$, for radially expanding 
solutions the quantity $R^2/G^2$ remains bounded, such that ${\cal D}$ does 
not reverse sign in this case, cf. Eq. (A.4). Conversely if the flow tends 
to be collimated or becomes radial very slowly the above quantity rapidly  
increases with $R$ for $R \gg 1$, leading to the appearance of this new 
singularity [${\cal D}=0$ in Eq. (A.4)]. This explains the emergence of the 
second critical point in such cases.

For a given range of $\kappa$-values, $[\kappa_1, \kappa_2]$, the two 
unphysical families which flare with $F_{\infty}=-2$ separate at a finite 
distance. In principle the only possible solution would be the critical one 
that crosses this new X-type critical point. However, it turns out that this
critical solution always shows downwind of the position of the critical point 
oscillations of increasing amplitude and eventually  terminates in a loop at a 
finite distance. The termination position gets closer to the critical point by 
further increasing $\kappa$ to less negative values and/or increasing 
$\epsilon$. This type of solutions may be physically unacceptable since they 
do not extend up to infinity, unless they are terminated by a shock, with a  
positive pressure (we discuss this point in more detail in the next paragraph).
In this case, this extra critical point seems to be the first of two or more
fast critical transitions which seem to appear in the \cite{WD67}
1-D solution topologies and also in the 2-D analysis of \cite{HN89}. 
Similarly to those examples, the solutions loop back rather sharply, returning 
upstream (Fig. \ref{f3}).

If the efficiency of the magnetic rotator increases further and/or $\kappa$ 
gets larger than $\kappa_2$, then a new family of cylindrically collimated 
solutions enters the picture. Topologically, they appear once the turning 
point of the terminated solutions reaches the X-type critical point. They have
the typical properties of those from an {\bf EMR} with 
$\kappa > 0$ (see STT02). The critical solution itself changes as it becomes 
also cylindrically collimated. 

The various families of solutions which exist in that case are as follows. 
First, with rather high pressures we have solutions which still flare with 
$F_\infty=-2$. Second, with rather low pressures we have the solutions which 
loop back. And finally, in between those two families of solutions, for a 
given intermediate range of $\Pi_*$, we have the third family of cylindrically
collimated ones. 

One member of this third family of solutions crosses the X-type singularity 
which is still present at the border between cylindrical and looping solutions.
The other members of this third family of solutions are noncritical. 
The critical solution is analogous to  the limiting solution of STT02 for 
$\kappa>0$.

Furthermore we know from the asymptotic analysis that different branches of 
solutions are present, corresponding to centrifugally or pressure supported 
flows. The present numerical results show that only the configuration with the
smallest transversal radius can be attained by the jet, which is supported by 
the centrifugal force. Thus, as predicted by the asymptotic analysis performed
in STT99,  all cylindrically collimated solutions including the critical ones 
have almost the same asymptotic behaviour. In other words the pressure plays a 
minor role in achieving the asymptotic configuration of these solutions.

\section{Properties of the critical solutions\label{sec5}}

\subsection{Cylindrically collimated solutions}

An interesting novel feature in the present cylindrically collimated solutions 
is the appearance of two X-type critical points within the flow domain, in 
addition to the Alfv\'en critical point. The only other known case where a 
unique steady MHD outflow solution is filtered by three critical points is the 
case of a radially self-similar solution (Vlahakis et al., 2000, 
\cite{FerreiraCasse04}). 
In general, at such critical points the bulk flow speed equals to one 
of the characteristic speeds in the problem. Hence, it is of physical interest 
to associate the  flow speeds at these critical X-type points to some  
characteristic MHD speeds. In that connection, we first note that the 
present solutions posses the symmetries of meridional self-similarity 
and axial symmetry.  Thus, in spherical coordinates 
(r, $\theta$, $\varphi$), the self-similarity direction is $\hat \theta$ and 
the axisymmetry direction is $\hat \varphi$. Therefore, a wave that preserves 
those two symmetries should propagate along the $\hat r$-direction in the 
meridional plane. 
{\it First}, the incompressible Alfv\'en mode propagates along the magnetic 
field $(\vec{B})$ with velocity $V_a$ and in the direction $\hat r$ of the 
poloidal plane with a phase speed 
$V_{a,r} = {\vec{B}}\cdot \hat r/\sqrt{4\pi \rho}$. 
Thus, at the Alfv\'en point we should have $M=1$. 
And {\it second}, the compressible slow/fast MHD modes propagate in the 
direction $\hat r$ with a phase speed 
$V_{\rm X} \equiv V_{slow, r}$, or,  $V_{\rm X} \equiv V_{fast, r}$ 
which satisfy the quartic
\begin{equation}
V_{\rm X}^4 - V_{\rm X}^2(V_a^2 +C_s^2) + C_s^2V^2_{a,r} = 0
\,.
\label{Eq19}
\end{equation}
Hence, when the above equation is satisfied the governing Eqs. (\ref{Eq1}), 
(\ref{Eq2}) and Eq. (\ref{Eq3}) have X-type singularities and 
$V_{\rm X} = V_p \cdot \hat r$. 

On the other hand, it is well known that in the MHD flow system there exist 
two hyperbolic regimes wherein characteristics exist: the inner, which is 
bounded by the cusp and the slow magnetosonic surfaces and the outer 
extending downstream of the fast magnetosonic point. Within each of those two 
hyperbolic regimes, there exists one limiting characteristic or separatrix 
surface: the slow magneto-acoustic separatrix surface (SMSS) inside the inner 
hyperbolic regime and the fast magneto-acoustic separatrix surface (FMSS) 
inside the outer hyperbolic regime (Bogovalov, 1994, Tsinganos et al., 1996).  
The true critical points are precisely found on these two separatrices. 
For example, in the case presented in Figs. 4, the SMSS is at $R= 0.751$ 
while  the FMSS is located at $R=4150$.  
 
In the underpressured solutions studied in STT02 we have found only the X-type 
critical point inside the inner hyperbolic regime wherein the radial outflow 
speed is  $V_r = V_{slow, r}$ (ST94). Now, in the regime of cylindrically 
collimated solutions (domain with stars in Fig. \ref{f5}), there exists a 
unique critical solution that also crosses the second critical point. This 
solution always has negative values of $\Pi(R)$ asymptotically.  Thus, in 
order to have positive values of the total pressure everywhere in the flow, we
have to adjust $P_o$ to some positive value (Eq. \ref{pressure}). The closer 
to zero is $\kappa$ the more negative becomes the  function $\Pi$ and the 
larger is the minimum value needed for $P_o$.

The fast magnetosonic nature of this second critical point can be analysed by 
drawing the characteristics in the vicinity of this separatrix critical 
surface, provided that we are able to define there the sound speed. 
Alhough the sound speed is ill-defined in our model, we can nevertheless 
deduce its value at the critical surface by following the steps presented in 
\cite{Tsinganosetal96},
\begin{equation}
C_s^2 = {\partial P \over \partial \rho}\Big|_{(R, \alpha)}
\nonumber
\end{equation}
\begin{equation}
=-{V_\star^2\over 2}
  {\partial \Pi (M^2, R)\over \partial M^2}\Big|_{(R, \alpha) }
M^4{1+\kappa\alpha\over 1+\delta\alpha } +H(R,\alpha)\,, 
\end{equation}
with $H(R,\alpha)$ a function which becomes zero at the critical surface. 
The condition at the critical point is equivalent to setting equal to zero 
the denominator [Eq. \ref{Eq4}] of Eqs. (\ref{Eq2}) and (\ref{Eq3}), 
\begin{equation}
(M_a^2-1)\left( 1+\kappa {R^2\over G^2} \right)
          + {F^2\over 4} + R^2\lambda^2{(1-G^2)^2\over (1-M_a^2)^2} =0
\,.
\end{equation}
Together with the previous definition of the sound speed and Eq. (\ref{Eq1}), 
this can then be put into the form of Eq. (\ref{Eq19}) with $V_X=V_r$. 

\begin{figure*}
%\vspace{11.0 cm}
\centerline{
\psfig{figure=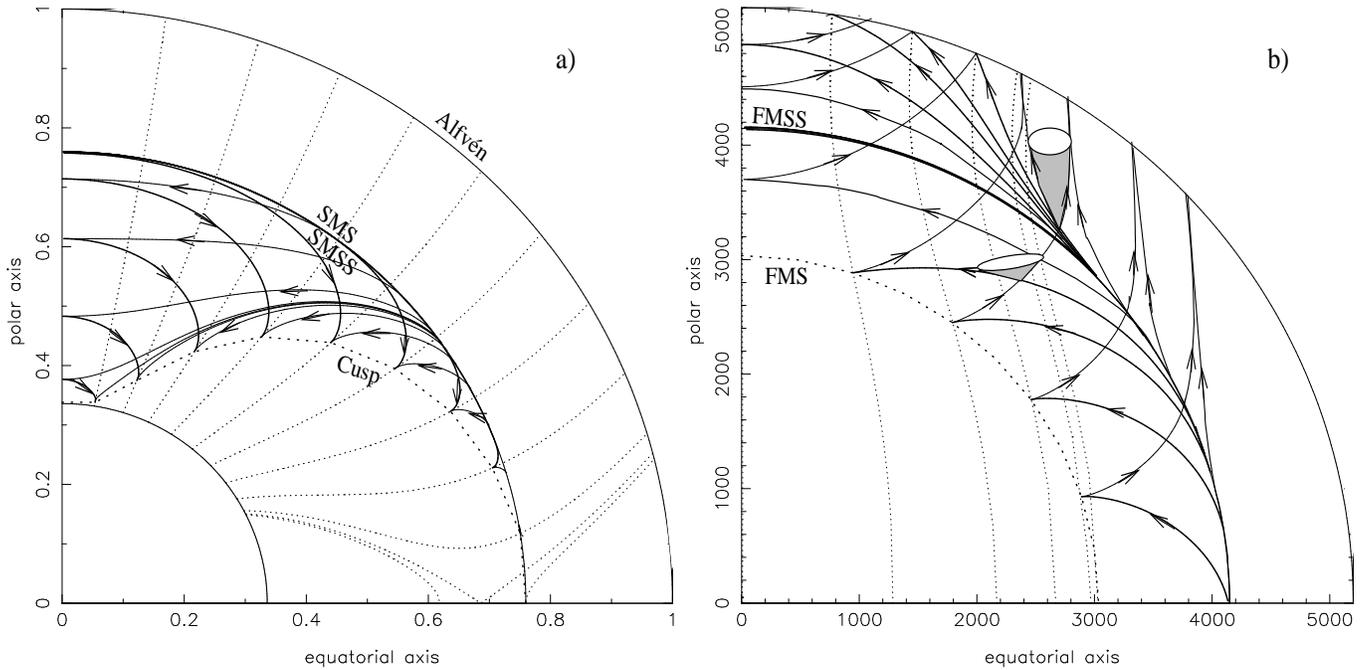,width=\textwidth,height=9cm,angle=0}}
\caption{Slopes of the two families of characteristics of the collimated
critical 
solution of Figs. \ref{f4} in each of the two hyperbolic regimes of the 
problem.  In a) the slow magnetoacoustic separatrix surface (SMSS) is at 
$R$=0.751 
just before the slow magnetoacoustic surface (SMS). In b) the fast 
magnetoacoustic separatrix surface (FMSS) is at $R$=4158 above the 
fast magnetoacoustic surface (FMS)  at about $R = 3000$. Arrows 
indicate the direction of 
MHD signal propagation while two Mach cones above and below the FMSS are also 
shown.
\label{f6}}
\end{figure*}

Moreover assuming that  $H(R)=0$ in all space we can calculate the slopes of 
the two characteritics in the regions where the equations are hyperbolic. The 
results are displayed on Figs. \ref{f6} for the cylindrically collimated 
solution of Figs. \ref{f4}. We show that there is an inner hyperbolic domain 
bounded downstream by the slow magnetosonic transition as well as an 
outer hyperbolic domain bounded upstream by the fast magnetosonic surface. 
As in \cite{Tsinganosetal96}, the transition from hyperbolic to elliptic in 
Fig. \ref{f6}a  is very close to the critical 
transition SMSS such that it is impossible to distinguish between the two 
without zooming closer than what is done in this figure. Instead in Fig. 
\ref{f6}b, the FMSS is clearly distinguishable from the FMS.
Thus, in both cases the magneto-acoustic separatrices clearly differ 
from the corresponding magneto-acoustic transitions in the same way 
the ergosphere differs from the event horizon of a rotating black hole 
(see \cite{Sautyetal02}). 

\subsection{Terminated solutions}

We have seen that the main feature of the terminated solutions is the onset, 
downstream of the FMSS, of oscillations with growing amplitude.
This is likely to be related to some instability that does affect the 
configuration of the outflow which thus cannot attain any steady configuration.
This is confirmed by the perturbative analysis in STT99: for this set of 
parameters the asymptotic solutions show a turning point (see Fig. \ref{f1},
left panel, and Figs. 1 in STT99) and in this region the typical oscillations 
of the asymptotic streamlines are not present, namely their wavelengths become 
imaginary (see Fig. \ref{f1}, right panel, and Fig. 8,  Eqs. 5.4 - 5.5 in 
STT99).

The cylindrical solutions with the second critical point are extended to 
infinity because there is some freedom on $\Pi_\infty$. For the terminated 
solutions, 
cylindrical asymptots are forbidden because the cylindrical regime is unstable
(infinite wavelength of the oscillations). The only remaining solution is for 
the lines to become radial. However, by crossing the second critical point the
value of $\Pi$ is fixed, so there is little chance that the solution becomes 
radial { or paraboloidal}
because the pressure does not vanish asymptotically. By plotting 
the forces across the lines, we see that in fact at the turning point the 
pressure gradient dominates. It is balanced only by the curvature force of the
poloidal velocity which creates the turning point and the termination of the 
solution. 

\subsection{Radial solutions}

In outflows which are launched dominantly thermally (cf. Fig. 2c), the 
streamlines start 
with a  basically radial shape close to their base. Further away,
if the outflow  is strongly overpressured ($\kappa < \kappa_1$), the 
asymptotics remains radial. Hence, there is no drastic change in the 
geometry of the streamlines. On the other hand, in mildly overpressured 
ouflows ($\kappa > \kappa_2$),  the asymptotics changes to cylindrical 
(in the intermediate case  $\kappa_1 < \kappa < \kappa_2$ the oscillations are 
so strong that the solution terminates). Thus, for $\kappa > \kappa_2$, there 
exists a transition region  between the base and the asymptotic regime wherein 
the cylindrical geometry is 
finally obtained after the basal roughly radial regime. During that transition 
phase the outflow naturally passes from a stage of oscillations in its radius, 
Mach number and other physical parameters. These oscillations can be 
understood 
as the result of the interplay of the pinching magnetic tension force and the 
resulting reaction by the flow (conservation of angular momentum).   
In consistency with the previous explanation, a conspicuous feature appearing 
in the asymptotically radially solutions is the lack of any oscillation of the 
poloidal streamlines, a result confirming the physical origin of these  
oscillations. 
   
In that connection, theoretical arguments and various analytical self-similar 
solutions have shown that a notable common feature of all self-consistent, 
self-similar MHD solutions which become finally cylindrically collimated is 
that the outflow passes from a stage of oscillations in its physical 
parameters (\cite{VT97}). 
Such oscillatory behavior of collimated outflows is not restricted to the few 
specific models examined so far, but instead it seems to be a rather general 
physical property of an MHD outflow which starts noncylindrically before it 
reaches collimation.  Note that the same feature of oscillations has been also 
found in non-self-similar simulations of outflows which start radially before 
the magnetic tension converts them to a cylindrical shape 
(\cite{Tsinganosetal03}).

\section{Conclusions\label{sec7}}

\subsection{Summary of results \label{subsec7.1}}

In this paper we continued the analysis of Meridional Self-Similar Models 
(MSSM, hereafter) by confining our attention to the 
study of outflows with a density increasing away from the axis and towards the 
surrounding streamlines [cf. Eq.  (\ref{density}) with $\delta >0$] 
and with a pressure decreasing from the axis [cf. Eq.  (\ref{pressure}) 
with $\kappa <0$]. 
In such {\it overpressured} outflows with a central dip in the density 
distribution, the temperature is strongly peaked at the axis relatively to 
the surrounding regions, more than in the {\it underpressured} 
outflows studied in STT02.

We have been able to construct solutions describing outflows starting 
subsonically and subAlfv\'enically from the central gravitating source and its
surrounding accretion disk and, after crossing the MHD critical points, 
reaching high values of the Alfv\'en Mach number. 

In terms of asymptotic profiles three broad types of solutions are found:  
\begin{itemize}
\item
(a) Collimated jet-type outflows from {\bf EMR} where the outflow is confined 
by the magnetic hoop stress, provided that they are not too overpressured  
($\kappa$ not too negative, i.e., $\kappa >\kappa_2$). 
Among those solutions, a unique one crosses three critical points: the SMSS, 
the Alfv\'en and the FMSS. We analysed those solutions and have 
shown that the separatrices indeed correspond to the three familiar 
separatrices in MHD wind theory. This class of critical solutions exhibits at
large distances oscillations between over- and  under-pressured flow (Fig.
\ref{f4}). 
\item
(b) Radially expanding wind-type outflows, analogous to the solar wind, 
for all  {\bf IMR} [cf. Eq.  (\ref{epsilon}) with $\epsilon <0$] or strongly 
overpressured sources from  {\bf EMR} ($\epsilon >0$, $\kappa <\kappa_1  $). 
Those solutions cross the slow and the Alfv\'en point and the initial pressure 
is fixed by the outer boundary condition that the terminal pressure should 
become zero. They do not show any intermediate topology of a third critical 
point. 
\item
(c) Terminated solutions.   
Such solutions cross again the three MHD separatrices. The onset of the 
increasing amplitude oscillations and the termination of the solutions can be 
understood because cylindrical asymptotics were shown to be unstable. 
As for disk wind solutions crossing all 
critical points, shown in \cite{Vlahakisetal00}, such terminated solutions can 
support terminal shocks. This is the opposite of the solutions of refocalizing 
disk wind usually used in models (e.g. \cite{FerroFontanDeCastro03}).
Nevertheless, our terminated solutions are usually having very high
temperature in the far regime and thus they are unphysical.
\end{itemize}

{ 
Although from the asymptotic analysis presented in STT99 we could not 
exclude the existence of paraboloidal asymptotics, we could not find 
numerically any such solution except in the limiting case 
$\epsilon=\kappa=0$ (ST94). 
We conjecture that these paraboloidal solutions could be found in principle
in the transition region between radial and cylindrical asymptotics.
However steady equilibrium is impossible in this region and
the solutions terminate before reaching their asymptotic regime. }

\subsection{Astrophysical implications\label{subsec6.2}}

{ Possibly related to the previous discussion on the lack of paraboloidal 
solutions,}
we  note that, conversely to underpressured jets studied in STT02, 
the transition of collimated jets to uncollimated winds is not continuous in 
the parametric space showing a gap where stationnary solutions do not exist. 
We are tempted to conjecture, then, for some jets with strong transient
events, such as violent outbursts, the following scenario:  if the outflow 
configuration is at the 
interface of the regime with collimated and non collimated solutions, an 
outburst 
could be associated with the flip over between the two different classes of 
asymptotically collimated and radial solutions.

One of the major outstanding questions in astrophysical jets research 
is how they are generated. Take for example the case of the closer and 
thus better resolved jets associated with young stellar objects. 
In that connection one may say that through a combination of observations 
and numerical simulations we do know several details about the propagation 
of these jets in the parent cloud and their interaction with their 
environment but we know relatively fewer details about their generation 
at the "central engine" (cf. \cite{Hartigan03}, \cite{RayBacciotti03}). 
At the same time various studies have shown that a high degree of 
collimation is already achieved very close to the source, namely at 10 
to 20 AU (\cite{Woitasetal02}).  

In that content, two wide classes of models are available today to study 
analytically the launching and eventual collimation of MHD outflows 
(\cite{VT98});  {\it first}, the family of the so-called 
radially self-similar models (RSSM) which have as their prototype the 
Blandford \& Payne (1982) model and {\it second} the family of the 
so-called meridionally self-similar models (MSSM) which have as their 
prototype the Sauty \& Tsinganos (1994) model 
and which we have explored in this series of papers. 
A third class of models 
are the so-called X-wind models where mass loss originates at a fan 
of concentrated magnetic flux in the inner disk radius (\cite{Shuetal94}, \cite{Shangetal02}).

In an analytical MHD treatment of the problem of outflow launching and 
subsequent collimation, in the RSSM the driving force and collimation 
mechanism are basically {\it magnetic}. Among the limitations of 
the RSSM is  
that they are invalid close to the jet axis where they have singularities and 
are thus more appropriate to describe disk-winds which collimate within 
several  AU from the star (\cite{Ferreira97}; henceforth F97). 
Also, after several 
Alfv\'en radii when the streamlines reach their maximum cylindrical radius, 
they slowly refocus towards the axis and the solutions terminate. Another 
difficulty of the cold plasma RSSM is that they predict rather large terminal 
speeds and too low densities and ionisation fractions and do not accomodate 
some efficient heating mechanism which is needed in order to explain the 
observed emission (\cite{Dougadosetal03}). 
However, Casse \& Ferreira (2000) have shown that by including a hot corona 
above 
the disk it would help to increase the mass loss rate and thus the terminal 
densities. 
In any case, the RSSM deal consistently with the accretion 
ejection problem and have been recently used with some success to compare 
observed jet widths and collimation scales in several T Tauri microjets for 
which we currently have the corresponding observations 
(Dougados et al., 2000, Pesenti et al., 2003).

On the other hand, in the MSSM for jet acceleration and collimation 
the driving force is a combination of thermal and magnetocentrifugal 
terms while collimation can be also achieved by a combination of pressure 
gradients and magnetic tension forces. In the MSSM however, if the 
source region of the outflow is restricted to be only the stellar base, the 
resulting mass loss is unrealistically low, unless it includes the
inner part of the disk.  
Nevertheless, an interesting fact is that observations clearly show 
that jets extend to relatively great distances (100 to 1000 AU) from the 
protostar where the observation of forbidden line emission means that the 
jets are still warm/hot at such large distances. 
But then if jets are launched from small regions and also expand, they 
should cool adiabatically. The question arises then on how do these jets 
remain hot at such distances from the protostar. Clearly a heating mechanism 
is needed. Hence, observations seem to suggest that thermal gradients, which 
may originate in a stellar or an accretion heated disk-corona, play an 
important role in accelerating the flow (Dougados, 2003). Such a heating 
is a basic ingredient of the MSSM and may thus explain the puzzle. In addition 
steady MSSM stay tightly collimated to unlimited distances from the source 
without a need to refocus towards their axis along which they are valid 
without any 
singularity. Furthermore, it has been suggested that HH jets may be the 
progenitors of the (uncollimated) solar wind outflow, a form in which jets 
eventually evolve after the star looses angular momentum becoming an 
inefficient magnetic rotator (ST94, STT99, STT02 and this paper).  

However, recent numerical simulations of magnetocentrifugally collimated 
outflows
from a rotating central object and/or a Keplerian accretion disk have shown 
that relatively low mass and magnetic fluxes reside in the produced jet as 
compared to the surrounding wind (Tsinganos \& Bogovalov, 2002, 
\cite{Mattetal03}). { This is also the case in the solutions presented
in ST94 and STT02. In ST94 it is pointed that a significant fraction of the 
total mass loss rate of the jet is originating in the disk.}
Observations however indicate that in jets from young stellar objects, the 
collimated outflow carries higher fluxes than these studies predict. 
As a solution to this problem it has been proposed that jets may be described 
as a two-component system composed of an outflow originating at a 
central object which is surrounded by a disk-wind (ST94, \cite{Koideetal98}, 
STT02, Tsinganos \& Bogovalov, 2002). 
In that respect, \cite{Hartiganetal95} have identified an outer low velocity 
component (LVC) with velocities in the range of 10 to 50 km s$^{-1}$ along 
with a high velocity component (HVC) with radial velocities of a few hundred 
km s$^{-1}$. 
According to Kwan \& Tademaru (1995) the LVC is probably a low-velocity 
disk-wind that encompasses the jet. This view was confirmed later by 
\cite{Bacciottietal02} 
using HST/STI data to investigate the velocity structure of the DG Tau jet.
They found that the kinematics follows an "onion-like" structure with HVC 
closer to the jet axis and a LVC spread out wider. 

{ Double component jets are also clearly appearing in time dependent 
simulations of  jets around black holes and stars (e.g. \cite{Kudohetal98},
\cite{Koideetal98}). 
The magnetic field does not penetrate the
black hole magnetosphere, thus the inner plasma is compressed and a pressure 
driven outflow develops. The surrounding wind is centrifugally driven 
from the disk and magnetically collimated. However the main difference
with double jet component simulations and the present analytical solutions
is that the first describe dense core jets while the second model
hollow jets. At this point note that both analytical (\cite{Hanaszetal00}) 
and numerical (\cite{Kudohetal02}) studies of the flow stability tend
to show that the inner flow is probably more stable than the Keplerian
outer part and this is particularly true in the case of hollow jets.
Further comparison of simulations and our analytical modelling is however 
difficult because the boundary conditions generally used in disk winds are
rather different from the self-similar assumptions used here.
}

In accordance with the above theoretical and observational difficulties 
encountered by the single-component models, we propose that 
jets may indeed be described as a two-component outflow system.   
The model presented in this paper and in STT02 is by itself a double 
component jet structure for cylindrically collimated solutions. 
One part of the jet comes from the star itself. The other comes from the inner
boundary of the disk which is connected with the stellar magnetosphere. Such 
double structure can be applied directly to model jets from T Tauri stars
with  low mass accretion rate, like  RY Tau for instance.
However for T Tauri stars having higher mass accretion rate, like DG Tau, a 
consistent model would be one with 
an inner outflow described by a ST94-type MSSM, surrounded 
by a wider 
%CS ok or not?
F97-type RSSM disk-wind part. 
For example, in the present MSSM the velocity is peaked at the axis and the 
degree of collimation increases with velocity, as observed. 

A supporting evidence for the previous scenarios comes from recent 
findings on the rotation of jets from T Tauri stars with high accretion rate,  
by using either 
near-infrared long-slit spectroscopy in a series of distant knots 
(several 1000 AU away), or, by using HST/STIS observations much closer 
(within 100 AU) (Bacciotti et al., 2002, \cite{Coffeyetal03}).   
The remarkable fact is that the magnitude of the only recently inferred 
toroidal velocity in the jets (5 - 15 km s$^{-1}$ at distances of 20 to 30 AU 
from 
the flow axis and at around 100 AU from the plane of the disk) is precisely 
what some time ago was already inferred from MHD jet launching models
(see for example,  \cite{TT91}, Fig. \ref{f6},  and Tsinganos et al. 1992, 
Figs. 2,3 for a MSSM; or, Vlahakis et al., 2000, Figs. 5,7 for a RSSM). 
It should be interesting then to further discuss 
new observations in the context of the two analytical available models (MSSM 
and RSSM).

\begin{acknowledgements}
We thank T. Kudoh for careful reading of the manuscript and his useful 
comments.
E.T.  acknowledges financial support from the Observatoire de Paris,
from the Conferenza dei Rettori delle Universit\`a Italiane (program Galileo)
and from the Italian Ministery of Education (MIUR). 
C.S. and K.T. acknowledge financial support from the French Foreign Office 
and the  Greek General Secretariat for Research and Technology 
(Program Platon and Galileo).  K.T. acknowledges partial support from the 
European Research and Training Networks PLATON (HPRN-CT-2000-00153) and 
ENIGMA (HPRN-CT-2001-0032). 
\end{acknowledgements}

\appendix

\section{MHD Equations for meridionally self-similar flows}\label{A}

Under the assumptions of axisymmetry and meridional self-similarity, the 
MHD equations reduce to the following three ordinary differential equations 
for $\Pi(R)$, $M^2(R)$ and $F(R)$ :
\begin{equation}\label{Eq1}
{\hbox {d} \Pi \over \hbox {d} R}=
- {2 \over G^4 }
   \left[ {\hbox{d} M^2 \over \hbox{d} R} + {M^2 \over R^2} (F-2) \right]
-  {\nu^2 \over M^2 R^2 }
\,,
\end{equation}
\begin{equation}\label{Eq2}
{\hbox {d} F(R) \over \hbox {d} R}={{{\cal N}_F(R,G,F,M^2,\Pi; \kappa, \delta,
\nu, \lambda)} \over \     {R \, {\cal D}(R,G,F,M^2; \kappa, \lambda)}}
\,,
\end{equation}
\begin{equation}\label{Eq3}
{\hbox {d} M^2(R) \over \hbox {d} R}={{{\cal N}_M(R,G,F,M^2,\Pi;
\kappa, \delta, \nu, \lambda)} \over \     {R \, {\cal D}(R,G,F,M^2; \kappa,
\lambda)}}
\,,
\end{equation}
where we have defined:
\begin{equation}\label{Eq4}
{\cal D}= (M^2-1)\left( 1+\kappa {R^2\over G^2} \right)
          + {F^2\over 4} + R^2\lambda^2{N_B^2\over D^2}
\,,
\end{equation}
\begin{eqnarray}\label{Eq5}
{\cal N}_F = -(\delta-\kappa)\nu^2 {R G^2\over 2 M^2}F
\nonumber\\
+ \left[{2\kappa \Pi G^2 R^2} + (F+1)(F-2) \right]
\times
\nonumber\\
\times
\left(1+\kappa {R^2\over G^2} - {F^2\over 4}
                       -R^2\lambda^2{N_B^2\over D^3} \right)
\nonumber\\
+{M^2F\over4}(F-2)\left(F+2+2\kappa{R^2\over G^2}
             +2R^2\lambda^2{N_B^2\over D^3} \right)
\nonumber\\
-\lambda^2 R^2 F(F-2){N_B\over D^2}
\nonumber\\
+\lambda^2 R^2 \left(1+\kappa {R^2\over G^2} -R^2\lambda^2{N_B^2\over D^3}
                     - {F\over 2} \right)
\nonumber\\
\left( 4{N_B^2\over D^2}-{2\over  M^2}{N_V^2\over D^2}\right) \,,
%\eqno (\hbox {B.3b})&
\end{eqnarray}
\begin{eqnarray}\label{Eq6}
{\cal N}_M= (\delta-\kappa)\nu^2 {R G^2\over 2 M^2}(M^2-1)
\nonumber\\
+ \kappa \Pi R^2 G^2 M^2{F\over 2}
-{M^4\over 4}(F-2)(4\kappa {R^2\over G^2} +F+4)
\nonumber\\
+{M^2\over 8}(F-2)(8\kappa {R^2\over G^2} +F^2+4F+8)
\nonumber\\
- \lambda^2 R^2 (F-2){N_B\over D}
\nonumber\\
+\lambda^2 R^2 (2M^2+F-2)\left(
 {N_B^2\over D^2} -{1\over 2 M^2}{N_V^2\over D^2}\right)
\,.
%\eqno (\hbox {B.3c})
\end{eqnarray}
with
\begin{equation}\label{Eq7}
N_B = 1 - G^2,\;\;\;\;N_V=M^2-G^2,\;\;\;\;D=1-M^2
\,.
\end{equation}
The definitions of the various parameters is discussed in Sec. 2.

At the Alfv\'en radius, the slope
of $M^2(R=1)$ is $p= (2 - F_*)/ \tau$, where $\tau$ is a solution of the
third degree polynomial:
\begin{equation}\label{Eq8}
\tau^3 + 2 \tau^2 + \left( {{\kappa \Pi_*} \over {\lambda^2}} +
{{F^2_* - 4} \over {4 \lambda^2}} - 1 \right) \tau +
{{(F_*-2)F_*} \over {2 \lambda^2}} = 0
\,,
\end{equation}
and  the star indicates values at $R=1$ (for details see ST94).
\par\vfill\eject

% -------------------------------------------------
\end{document}